\documentclass[aps,prb,twocolumn,superscriptaddress,showpacs]{revtex4-1}

\usepackage{graphicx}
\usepackage{amsmath}
\usepackage{amssymb}
\usepackage{color}

\newcommand{\ket}[1]{\left|{#1}\right>}

\begin{document}


\title{Leakage-current lineshapes from inelastic cotunneling in the Pauli spin blockade regime}


\author{W. A. Coish}
\affiliation{Department of Physics, McGill University, Montr{\'e}al, Qu{\'e}bec H3A 2T8, Canada}
\author{F.~Qassemi}
\affiliation{Institute for Quantum Computing and Department of Physics and Astronomy, University of Waterloo, Waterloo, Ontario N2L 3G1, Canada}


\date{\today}

\begin{abstract}
We find the leakage current through a double quantum dot in the Pauli spin blockade regime accounting for inelastic (spin-flip) cotunneling processes.  Taking the energy-dependence of this spin-flip mechanism into account allows for an accurate description of the current as a function of applied magnetic fields, gate voltages, and an inter-dot tunnel coupling.  In the presence of an additional local dephasing process or nonuniform magnetic field, we obtain a simple closed-form analytical expression for the leakage current giving the full dependence on an applied magnetic field and energy detuning.  This work is important for understanding the nature of leakage, especially in systems where other spin-flip mechanisms (due, e.g., to hyperfine coupling to nuclear spins or spin-orbit coupling) are weak, including silicon and carbon-nanotube or graphene quantum dots.
\end{abstract}


\maketitle

\section{Introduction}

Spin-dependent current blockade effects have been observed in quantum dots \cite{Ono2002a,Pioro-Ladriere2003a} as well as in molecular\cite{Heersche2006a} and single-atom tansport.\cite{Loth2010a}  These effects are the basis of spin-to-charge conversion schemes, essential for measurements of spin coherence and relaxation \cite{Petta2005a,Petta2005b,Koppens2006a,Pioro-Ladriere2008a} as well as the accurate initialization and readout of spin states for spin-based quantum information processing.\cite{Loss1998a}  Blockade effects have further allowed the observation of intriguing slow periodic oscillations in current, dependent on nuclear spins.\cite{Ono2004a} A detailed microscopic understanding of how this blockade can be lifted is important to develop an accurate description of these effects and to point the way to generate a more robust blockade for the study of further spin-dependent phenomena. 

The Pauli spin blockade of current through a double quantum dot occurs when each of two quantum dots in series energetically favors a one-electron configuration (we will refer to this as the $(1,1)$ regime, where $(n,m)$ refers to $n$ electrons on the left dot and $m$ electrons in the right).  Restricting to only the lowest non-degenerate single-particle orbital state in each dot, there are four possible spin configurations in the $(1,1)$ subspace: one spin singlet and three spin triplets.  To generate sequential transport of electron charge from left to right, the double-dot must pass through the $(0,2)$ charge configuration, but due to the Pauli exclusion principle, the lowest-energy $(0,2)$-state is a spin singlet when only the lowest single-particle orbital state is accessible. An inter-dot tunnel coupling preserves the spin of the two-electron state and therefore couples only the $(1,1)$ singlet to the $(0,2)$ state.  After a small number of electrons has passed through the double-dot, eventually one of the spin-triplet states will be occupied by chance, leaving the double dot stuck in a  ``blocked'' configuration.  This blockade can be lifted either through the direct hybridization of singlet and triplet states with the addition of spin-non-conserving terms to the Hamiltonian (due, e.g., to the spin-orbit or hyperfine interactions), or through direct energy-conserving transitions between triplet and singlet levels.  In spite of this relatively simple explanation for the Pauli spin blockade, the situation is complicated by several possible microscopic mechanisms that may dominate in determining the leakage current depending on the material and device characteristics.  It is therefore important to understand precisely what influence each of the possible microscopic mechanisms may have on the overall leakage current in order to identify the most relevant mechanism and possibly to suppress it.

Most mechanisms that lift spin blockade are particular to the materials used to manufacture a double-dot device; the contact hyperfine interaction between electron and nuclear spins lifts the blockade in GaAs double dots,\cite{Ono2004a,Koppens2005a,Jouravlev2006a,Inarrea2007a} and a strong spin-orbit interaction plays the predominant role in lifting the blockade in InAs nanowire double dots.\cite{Pfund2007a,Danon2009a,Nadj-Perge2010a}  Both of these mechanisms can be suppressed by manufacturing double dots using silicon\cite{Liu2005a,Liu2008a,Shaji2008a,Lai2010a} or carbon-based\cite{Buitelaar2008a,Churchill2009a} materials, in which the majority isotope has no nuclear spin and the spin-orbit coupling strength $\propto Z^4$ is significantly weaker due to a smaller atomic number $Z$.  One blockade-lifting mechanism that is present in all double-dot devices, independent of the material composition, is exchange of spins with the leads through higher-order tunneling (cotunneling) processes.\cite{Fujisawa2003a,Liu2005a,Vorontsov2008a,Qassemi2009a}  By understanding and controlling these processes, one can accurately calibrate single-spin readout and improve on rapid spin preparation schemes.\cite{Qassemi2009a} Moreover, cotunneling processes have been shown to be significant in determining dynamic nuclear-spin polarizaiton processes, both in experiment\cite{Baugh2007a} and in theory,\cite{Baugh2007a,Rudner2007a} so a further understanding of cotunneling may allow for the preparation of a more highly-polarized nuclear-spin system.

In this article we derive analytic expressions for leakage-current lineshapes accounting for inelastic cotunneling processes in well-defined and generically accessible limits.  Some results of this analysis have recently been shown, experimentally, to be consistent with transport measurements on silicon double quantum dots,\cite{Lai2010a} and have proven useful in determining microscopic parameters associated with those devices.  A similar application of the results presented here to other material systems may shed light on, e.g., unusually broad lineshapes in the magnetic-field-dependent current through isotopically enriched $^{13}$C nanotube double dots.\cite{Churchill2009a}

Inelastic (spin-flip) cotunneling has been known as a significant spin-flip mechanism since early measurements of triplet-to-singlet decay in vertical double quantum dots,\cite{Fujisawa2002b} where the triplet-to-singlet decay rate was shown to be limited by inelastic cotunneling.\cite{Fujisawa2003a}  In the context of the Pauli spin blockade, spin-flip cotunneling rates have been calculated and compared to experimental data in the high-temperature regime where the associated transition rates between energy levels are independent of the energy-level spacing.\cite{Liu2005a,Vorontsov2008a} More recently, the consequence of the full energy dependence of these rates has been calculated\cite{Qassemi2009a} and verified in experiment.\cite{Lai2010a}  In this paper we apply and extend the analysis presented in refs. \onlinecite{Qassemi2009a} and \onlinecite{Lai2010a} to a broader range of parameters and provide a general and intuitive formalism for the calculation of leakage current through blockaded structures.  For simplicity, in specific calculations we neglect orbital/valley degeneracy in our treatment, which may be relevant for quantum dots made from graphene, carbon nanotubes, or silicon nanostructures and can lead, in general, to a more complicated spin-valley blockade.\cite{Palyi2009a,Palyi2010a} However, the general formalism we present can also be applied directly to systems with valley degeneracy and many of the results we present will be qualitatively unchanged in the presence of additional orbital degeneracies.

The remainder of this article is organized as follows: In Sec. \ref{sec:blockade-current} we present an intuitive and general procedure for the calculation of current through blockaded structures given a set of decay rates obtained from a microscopic calculation.  In Sec. \ref{sec:PSBCotunneling} we specialize to the case of a double quantum dot in the Pauli spin blockade regime.  We recall the calculation of sequential-tunneling and spin-flip cotunneling rates from ref. \onlinecite{Qassemi2009a} and apply the procedure of Sec. \ref{sec:blockade-current} to find leakage-current lineshapes as a function of an applied magnetic field and energy detuning (the energy difference between $(1,1)$ and $(0,2)$ charge states).  In the limit of a strong local spin dephasing mechanism or nonuniform magnetic field, we then obtain a single simple closed-form analytical expression giving a full two-dimensional map of the current as a function of detuning and magnetic field.   In Sec. \ref{sec:Conclusions} we conclude with a summary of the main results and a discussion of extensions and possible future work.

\begin{figure}
\vspace{5mm}
\includegraphics[width = 0.45\textwidth]{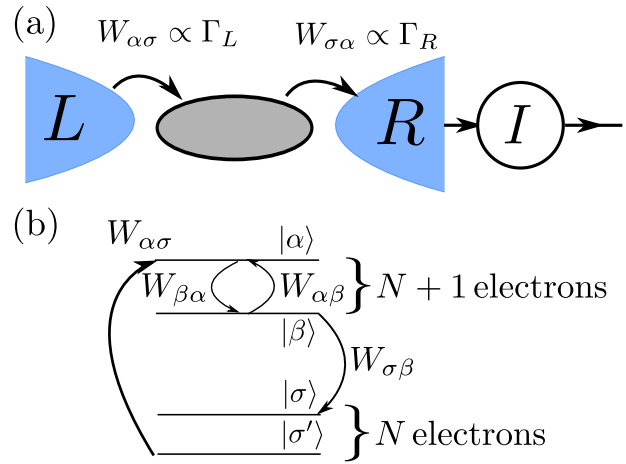}%
\caption{\label{fig:transport-fig} (Color online) (a) A typical transport scenario in the sequential-tunneling regime.  In the high-bias limit ($\delta\mu>|E_{\alpha}-E_{\sigma}|,T$, where $E_j$ is the energy of level $j$) electrons hop only in one direction: from the left lead to the dot at a rate $\propto \Gamma_L$, and from the dot to the right lead at a rate $\propto \Gamma_R$.  (b) The transport cycle in the  sequential tunneling regime is a loop that carries the dot from an $N$-electron state (labeled with $\{\sigma,\sigma'\}$) to an $N+1$-electron state (labeled here with Greek letters $\{\alpha,\beta\}$).  A current blockade will be set up if one or some of the accessible $N+1$-electron states $\alpha$ have negligible transition rates $W_{\sigma\alpha}\simeq 0$, in which case the current may be limited by rates $W_{\beta\alpha}$ inducing transitions within the $N+1$-electron subspace.}
\end{figure}

\section{Leakage current through blockaded systems}
\label{sec:blockade-current}
In this section we establish a generic theory of the current through blockaded nanostructures, incorporating internal transition rates between levels that can lead to a lifting of the blockade.  The formalism developed in this section will be used in later sections to derive simple analytical formulas for the dependence of the current on various parameters in the specific case of the Pauli spin blockade of transport through a double quantum dot, accounting for inelastic cotunneling processes.  Throughout this paper, we work in units where $g\mu_\mathrm{B}=e=k_\mathrm{B}=\hbar=1$, with electron $g$-factor $g$, Bohr magneton $\mu_\mathrm{B}$, electron charge $e$, Boltzmann's constant $k_\mathrm{B}$ and Planck's constant $\hbar$.

We consider a nanoscale system weakly tunnel coupled to leads, set at chemical potentials $\mu_l$ and with tunneling rates $\Gamma_{l}$ ($l=\mathrm{L,R}$ for the tunneling rate between the system and the left and right leads, respectively, see Fig. \ref{fig:transport-fig}(a)).  In the sequential-tunneling regime, electron charge is transported from left to right through energy-conserving transitions between $N$-electron states of the system, denoted $\sigma$, and $N+1$-electron states, denoted $\alpha$.  These transitions are associated with the exchange of an electron with one of the leads (Fig. \ref{fig:transport-fig}(b)).  Finally, we consider the simplifying limit of a large bias $\delta\mu = \mu_L-\mu_R>0$:
\begin{eqnarray}
 |\mu_L-\mu_{\alpha\sigma}|>T,\label{eq:high-bias1}\\
 |\mu_R-\mu_{\sigma\alpha}|>T,\label{eq:high-bias2}
\end{eqnarray}
with chemical potential $\mu_{\alpha\sigma}=E_\alpha-E_\sigma$ (where $E_i$ is the energy of isolated system level $i$ and Latin characters are taken to run over all system eigenstates, independent of the occupation number, i.e.: $i=\{\alpha,\sigma\}$).  In this limit, energy-conserving transitions that add one electron to the system $\sigma\to\alpha$ (with rate $W_{\alpha\sigma}$) necessarily involve the removal of an electron from the left lead and transitions that remove an electron from the system, $\alpha\to\sigma$ (with associated rate $W_{\sigma\alpha}$), involve the addition of an electron to the right lead (see Fig. \ref{fig:transport-fig}(a)).  Other processes that change the electron number are exponentially suppressed.

The sequential-tunneling current in the high-bias limit (defined by Eqs. \eqref{eq:high-bias1} and \eqref{eq:high-bias2}) is given simply by
\begin{equation}\label{eq:stcurrent}
 I = \sum_{\alpha\sigma} W_{\alpha\sigma}\bar{\rho}_{\sigma},
\end{equation}
where $\bar{\rho}_i$ solves the (stationary) Pauli master equation for the diagonal elements of the system density matrix:
\begin{equation}\label{eq:pauli-master-equation}
\overline{\dot{\rho}}_i = \sum_j W_{ij}\bar{\rho}_j-W_i\bar{\rho}_i = 0;\quad W_i=\sum_j W_{ji},
\end{equation}
where we define the average $\overline{\rho(t)} = \lim_{\tau\to \infty} \frac{1}{\tau}\int_0^\tau dt \rho(t)$. The diagonal elements of the stationary density matrix $\bar{\rho}_i$ must satisfy the normalization
\begin{equation}\label{eq:rhonorm}
 \sum_i \bar{\rho}_i=1.
\end{equation}
Use of the classical (Pauli) master equation to describe the diagonal elements of the system density matrix is strictly valid in the high-bias, weak-coupling limit $|\delta\mu|>\Gamma_{L,R}$, where coherences (off-diagonal elements with respect to the isolated system energy eigenbasis) decay to zero on a time scale $\sim 1/|\delta\mu|$ that is short compared to the tunneling time $\sim 1/\Gamma_l$. In Eq. \eqref{eq:stcurrent}, we have explicitly assumed that higher-order current-carrying cotunneling corrections $\propto \Gamma_L\Gamma_R$ are small relative to the sequential-tunneling terms $\propto \Gamma_l$.  In Sec. \ref{sec:PSBCotunneling}, we will account for cotunneling processes that do not carry current, $\propto \Gamma_l^2$, involving exchange with the same lead.  These processes typically dominate over the current-carrying cotunneling processes in the case of a double quantum dot considered in Sec. \ref{sec:PSBCotunneling}.

 Solving the linear system given by Eqs. \eqref{eq:pauli-master-equation} and \eqref{eq:rhonorm} for the stationary populations $\bar{\rho}_{\sigma}$ is sufficient to determine the current $I$ from the set of all rates $W_{ij}$.  However, it is physically intuitive to switch to new variables $k_i$, defined in terms of the current $I$, stationary populations $\bar{\rho}_i$ and the total escape rate from state $i$, $W_i = \sum_{j} W_{ji}$, as:
\begin{equation}\label{eq:kdefinition}
 k_i = \frac{W_i\bar{\rho}_i}{I} = \frac{\textrm{flux out of state $i$}}{\textrm{flux into all states $\alpha$}}.
\end{equation}
The quantities $k_i$ have a natural physical interpretation: $k_i$ is the number of times state $i$ will be visited, on average, per transport cycle ($\sigma\to\alpha\to\sigma'\to\ldots$, depicted in Fig. \ref{fig:transport-fig}(b)).  From the definition \eqref{eq:kdefinition} and normalization \eqref{eq:rhonorm}, we find the current
\begin{equation}\label{eq:current-k-formula}
 I = \left(\sum_i \frac{k_i}{W_i}\right)^{-1}.
\end{equation}
This formula can be understood directly in terms of the interpretation given above for the coefficients $k_i$.  The average time to leave state $i$ if it were occupied is $1/W_i$, while $k_i$ is the number of times (on average) that state $i$ is occupied in each transport cycle.  The ratio $k_i/W_i$ is therefore the average time spent in state $i$ per transport cycle and so the total average time per transport cycle (average time to transfer an elementary electron charge) is simply $\sum_i k_i/W_i$.  The inverse of this time is the rate at which charge is transfered from left to right, giving the current, Eq. \eqref{eq:current-k-formula}.

The coefficients $k_i$ can be found systematically in terms of the rates $W_{ij}$ using the identity
\begin{equation}\label{eq:ki-linear-system}    
k_i = \sum_j P_{ij}k_j;\quad\quad P_{ij} = W_{ij}/W_j,
\end{equation}
where the branching ratios, $P_{ij}$, give the probability for a transition to state $i$ conditioned on starting in state $j$.  Eq. \eqref{eq:ki-linear-system} follows directly from Eq. \eqref{eq:pauli-master-equation} and the definition $k_i \propto W_i \bar{\rho}_i$ (Eq. \eqref{eq:kdefinition}).  Solving the linear system given in Eq. \eqref{eq:ki-linear-system} and substituting the result into Eq. \eqref{eq:current-k-formula} is formally equivalent to solving for the populations $\bar{\rho}_i$ and substituting the result into Eq. \eqref{eq:stcurrent} for the current.  However, for the particular case of blockaded systems, we will find that Eq. \eqref{eq:ki-linear-system} lends itself better to approximation schemes and often the solution for the $k_i$ can be determined quickly on physical grounds without directly solving the linear system. 

\section{Inelastic cotunneling and the Pauli spin blocakde regime}\label{sec:PSBCotunneling}
Here we apply the formalism of Sec. \ref{sec:blockade-current} to perform an explicit microscopic calculation for the leakage current through a double quantum dot in the Pauli spin blockade regime, accounting for transition rates due to inelastic cotunneling processes.  Inelastic cotunneling is a second-order tunneling process associated with a change in energy of the isolated quantum-dot state with a compensating change in energy of the lead state.  Since the total energy is conserved, the energy of the combined dot-plus-leads system is, of course, unchanged in this process. 

The simple formalism derived in the previous section allows us to efficiently obtain closed-form analytical expressions for the leakage current in terms of all transition rates $W_{ij}$.  We begin by reviewing the calculation of transition rates due to sequential tunneling and inelastic cotunneling, presented in ref. \onlinecite{Qassemi2009a}.

\begin{figure}
\includegraphics[width=0.45\textwidth]{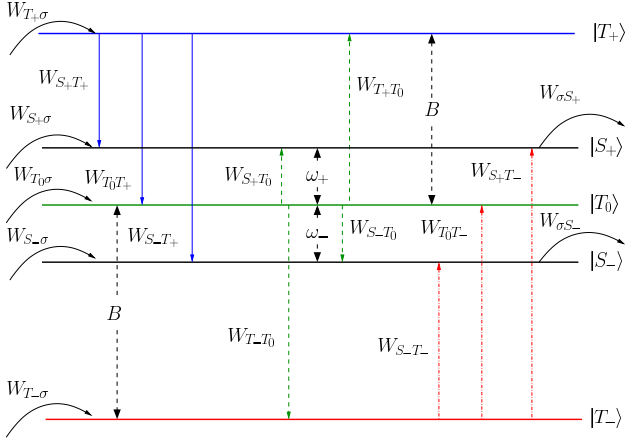}%
\caption{\label{fig:transitions} (Color online) Isolated double-quantum-dot energy eigenstates and transition rates between them.  The spin-polarized triplet states $\left|T_\pm\right>$ are split from the spin-unpolarized triplet $\left|T_0\right>$ by an applied magnetic field $B$ and the hybridized spin-singlet states $\left|S_\pm\right>$ are split from $\left|T_0\right>$ by the detuning-dependent quantities $\omega_\pm$, given in Eq. \eqref{eq:omega-pm}, below. }
\end{figure}

\subsection{Transition rates}

Our starting point is a Hamiltonian,
\begin{equation}
 H = H_0 + \sum_{l} H_{\mathrm{d}l},
\end{equation}
where $H_0$ gives the free Hamiltonian of the double quantum dot and Fermi-liquid leads.  In the subspace of $(1,1)$ and $(0,2)$ charge states, the isolated double-dot Hamiltonian eigenstates consist of two spin-singlets $\left|S_\pm\right>$, describing $(1,1)$ and $(0,2)$ charge states hybridized due to inter-dot tunneling $t$, as well as three spin triplets ($\left|T_\pm\right>,\left|T_0\right>$), see Fig. \ref{fig:transitions}.  Sequential transport further involves the $(0,1)$ charge state with spin $\sigma=\uparrow,\downarrow$, denoted: $\left|\sigma\right>$.  The Hamiltonian $H_{\mathrm{d}l}$ describes the coupling of lead $l$ to dot $l$ ($l=L,R$ for the left and right dot and lead, respectively), with tunneling amplitude $t_l$ (a full specification is given in the appendix, below). 

Transition rates are calculated via Fermi's golden rule:
\begin{equation}\label{eq:GoldenRule}
 W_{km} = 2\pi\sum_{if} \rho(i) \left|\left<fk\right|V\left|im\right>\right|^2\delta\left(\mathcal{E}_{fk}-\mathcal{E}_{im}\right),
\end{equation}
where $i,f$ label the initial and final states of the leads, respectively, $m,k$ label the initial and final states of the double dot, $\rho(i)$ describes a product of initial Fermi distributions in the leads, held at electron temperature $T$ and chemical potentials $\mu_l$, and $\mathcal{E}_{im},\mathcal{E}_{fk}$ give the total initial and final energies of both the double-dot and leads.  The effective perturbation $V$, accounting for transitions up to second order in $H_{\mathrm{d}l}$, is
\begin{equation}\label{eq:VDefinition}
 V = \sum_l H_{\mathrm{d}l}+\sum_{abgll'}\left|a\right>\left<b\right|\frac{\left<a\right|H_{\mathrm{d}l}\left|g\right>\left<g\right|H_{\mathrm{d}l'}\left|b\right>}{\mathcal{E}_a-\mathcal{E}_g},
\end{equation}
where the indices $a,b,g$ describe the collective state of the double dot and leads.

In the high-bias regime (Eqs. \eqref{eq:high-bias1} and \eqref{eq:high-bias2}), we find the sequential tunneling rates from Eqs. \ref{eq:GoldenRule} and \ref{eq:VDefinition}:
\begin{eqnarray}
 W_{\alpha\sigma} = \Gamma_{L}\sum_{\sigma'}|A_{L\sigma}^{\alpha\sigma}|^2\label{eq:Walphasigma},\\
 W_{\sigma\alpha} = \Gamma_{R}\sum_{\sigma'}|A_{R\sigma}^{\alpha\sigma}|^2\label{eq:Wsigmaalpha},
\end{eqnarray}
where the tunneling rates $\Gamma_l$ and transition matrix elements $A^{jj'}_{l\sigma}$ are, respectively,
\begin{eqnarray}
\Gamma_l &=& 2\pi\nu_l|t_l|^2,\\
A^{jj'}_{l\sigma} &=& \left<j\right|d^\dagger_{l\sigma}\left|j'\right>. 
\end{eqnarray}
Here, $\nu_l$ is the density of states per spin at the Fermi level in lead $l$.  We assume that both $\nu_l$ and $t_l$ are approximately energy-independent in our regime of interest.\footnote{The tunneling amplitudes $t_l$ will be approximately energy-independent when the bias is small compared to the height of the barrier coupling dot to lead. For a clean system, the density of states $\nu_l$ will be independent of energy as long as the bias is small compared to the Fermi energy $E_\mathrm{F}$.} The operator $d_{l\sigma}^\dagger$ creates an electron in single-particle orbital $l$ with spin $\sigma$.

The inelastic cotunneling rates $W_{\alpha\beta}$, arising from the second-order term in Eq. \eqref{eq:VDefinition} with $l=l'$, are given by\cite{Qassemi2009a}
\begin{equation}\label{eq:cotunneling-rates} 
W_{\alpha\beta}  =  2cTM_{\alpha\beta} F(\omega_{\beta\alpha}/T),
\end{equation}
with dimensionless prefactor $c$, matrix elements $M_{\alpha\beta}$, and energy-dependent factor $F(\omega/T)$ given  by
\begin{eqnarray}
 c &=& \frac{1}{\pi}\left[\left(\frac{\Gamma_L}{\Delta-\epsilon}\right)^2+\left(\frac{\Gamma_R}{\Delta+\epsilon-2U'-2\delta\mu}\right)^2\right],\label{eq:cdefinition}\\
M_{\alpha\beta} &=&\sum_{\sigma\sigma'\sigma^{''}} |A_{L\sigma}^{\alpha\sigma^{''}}|^2|A_{L\sigma'}^{\beta\sigma^{''}}|^2=\sum_{\sigma\sigma'j} |A_{R\sigma}^{j\alpha}|^2|A_{R\sigma'}^{j\beta}|^2,\label{eq:MCoefficient}\\
& & F(\omega/T) = \frac{\omega/T}{1-e^{-\omega/T}}.\label{eq:Fdefinition}
\end{eqnarray}
The coefficient $c\sim \Gamma_l^2$ in Eq. \eqref{eq:cdefinition} reflects the second-order nature of the cotunneling process and we have taken the convention $\mu_L=0$, $\mu_R=-\delta\mu$ with positive bias $\delta\mu>0$.  Here, the energy detuning $\epsilon = E_{(1,1)}-E_{(0,2)}$ measures the separation in energy between $(1,1)$ and $(0,2)$ charge configurations, and $\Delta = E_{(1,1)}-U$ sets the energy of the $(1,1)$ charge configuration with on-site charging energy $U$.  The transition matrix elements $M_{\alpha\beta}$ arise from processes involving spin exchange with the left lead (associated with virtual states in the $(0,1)$-subspace) or spin exchange with the right lead (associated with virtual states $\left|j\right>$ in the $(1,2)$-subspace).  The dominant energy and temperature dependence of the cotunneling rates is due to the function $F(\omega/T)$, which arises from an integral over Fermi functions $f_l(E) = 1/(e^{(E-\mu_l)/T}+1)$:
\begin{equation}
 F(\omega/T) = \frac{1}{T}\int_{-\infty}^\infty dE f_l(E)\left[1-f_l(E+\omega)\right].
\end{equation}
In our analysis, we have neglected resonant cotunneling contributions, which formally lead to a divergence in evaluating rates directly from Eq. \eqref{eq:VDefinition}.  However, these contributions can be systematically regularized\cite{Koenig1997a} and are suppressed exponentially in the high-bias limit considered here.  We have further neglected current-carrying cotunneling processes (those arising from the second-order term with $l\ne l'$ in Eq. \eqref{eq:VDefinition}).  We find that these processes are suppressed relative to the considered processes by at least a factor $\sim U'/U$, where $U'$ is the nearest-neighbor charging energy.\cite{Qassemi2009a}

Since the energy dependence of the rates $W_{\alpha\beta}$ will play an important role in the following analysis, it is useful to consider $F(\omega/T)$ in the limits of large positive and negative energy difference $\omega$ at low $T$: 
\begin{eqnarray}
 F(\omega/T) \simeq \frac{\omega}{T} \Theta(\omega/T),\quad\omega> T,\label{eq:F-positive-omega}\\
F(\omega/T) \simeq |\omega/T|e^{-|\omega/T|},\quad\omega<-T.\label{eq:F-negative-omega}
\end{eqnarray}
Eq. \eqref{eq:F-positive-omega} reflects the fact that the inelastic relaxation rates increase for large energy-level separation $\omega>T$, as the density of states of the environment increases and Eq. \eqref{eq:F-negative-omega} describes exponential suppression of excitation compared to relaxation processes, consistent with detailed balance.  

In the high-temperature limit, the inelastic cotunneling rates for both excitation and relaxation approach a constant, energy-independent value since, in this limit,
\begin{equation}
F(\omega/T) \simeq 1,\quad T>|\omega|.\label{eq:F-large-T}
\end{equation}
This high-$T$ limit has been explored in the context of Pauli spin blockade in previous works.\cite{Liu2005a,Vorontsov2008a}  In the present work, we are more concerned with the limits where the energy dependence described by Eqs. \eqref{eq:F-positive-omega} and \eqref{eq:F-negative-omega} is significant in determining the leakage current.\cite{Qassemi2009a}

\subsection{Leakage current: No local dephasing}\label{sec:no-dephasing}

Solving the linear system (Eq. \eqref{eq:ki-linear-system}) for $k_i$ with the rates given in Eqs. \eqref{eq:Walphasigma}, \eqref{eq:Wsigmaalpha}, and \eqref{eq:cotunneling-rates} immediately gives the current via the expression in Eq. \eqref{eq:current-k-formula}.  For any set of parameters, one can find the leakage current by solving the full linear system, giving a complex expression in general.  However, to understand the physical significance of the results, or in order to perform experimental fits to traces of leakage current vs. magnetic field or energy detuning, it is useful to derive simple analytical expressions, valid in experimentally relevant limits.  In this section, we derive expressions for the leakage current in the limit where there is no significant local dephasing mechanism, leading to decay rates that are comparable for all three spin-triplet states, $W_{T_+}\sim W_{T_-}\sim W_{T_0}$.  This limit applies to double dots in silicon or carbon when there is no magnetic field gradient.  In Sec. \ref{sec:strong-dephasing} below, we consider the opposite limit of a strong local spin dephasing mechanism or nonuniform magnetic field, leading to $W_{T_0}\gg W_{T_\pm}$, in which case the analysis simplifies considerably.

\subsubsection{B-field dependence (high-T limit)}
First we restrict ourselves to the dependence of the leakage current on an applied magnetic field $B$, which splits the spin-polarized triplet states $\left|T_\pm\right>$ from $\left|T_0\right>$ (see Fig. \ref{fig:transitions}).  We further consider the limit $\Gamma_{l}\gg W_{\alpha\beta}$  at zero detuning, $\epsilon=0$, in which the current is dominated by rates of escape from the three spin-triplet states $W_{T_a}\ll W_{S_\pm},W_{\sigma}$.  Noting that $k_{S_\pm},k_\sigma \sim O(1)$ in this limit since the singlets $\left|S_\pm\right>$ and one-electron states $\left|\sigma\right>$ are accessed at most once per transport cycle, Eq. \eqref{eq:current-k-formula} simplifies to
\begin{equation}\label{eq:current-eps-zero}
 I \simeq \left(\frac{k_{T_0}}{W_{T_0}}+\frac{k_{T_+}}{W_{T_+}}+\frac{k_{T_-}}{W_{T_-}}\right)^{-1},\quad \Gamma_{l}\gg W_{\alpha\beta}.
\end{equation}

\begin{figure}
\includegraphics[width=0.45\textwidth]{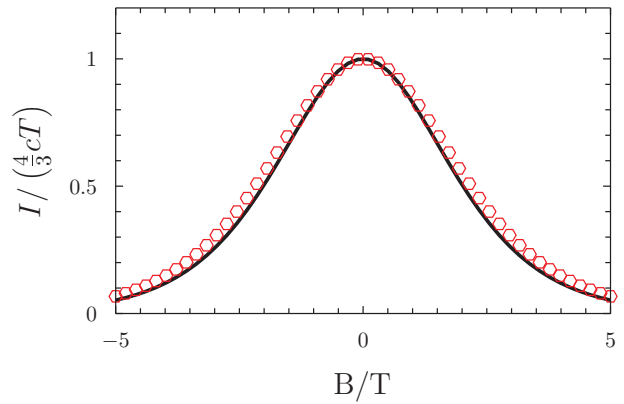}%
\caption{\label{fig:IvsB} (Color online) Current $I$ vs. magnetic field $B$ at zero detuning ($\epsilon=0$) from the full expression given in Eq. \eqref{eq:IvsBfull} (black solid line) and the approximation, Eq. \eqref{eq:IvsBapprox} (red circles).  This form is valid when the temperature $T$ is large compared to the tunnel coupling $t$ and when the direct dot-lead tunneling rates $\Gamma_l$ are large compared to inelastic cotunneling rates $W_{\alpha\beta}$, i.e. $T>t$ and $\Gamma_l>W_{\alpha\beta}$.}
\end{figure}

We obtain the relevant coefficients $k_\alpha$ from Eq. \eqref{eq:ki-linear-system}, with the cotunneling rates given in Eq. \eqref{eq:cotunneling-rates}.  We find $k_\sigma=1/2$, since both spin states, $\sigma=\uparrow,\downarrow$, are equally probable and one of the two is accessed in each transport cycle.  For an unpolarized source lead, we find the branching ratios $P_{T_+\uparrow}=P_{T_-\downarrow}=1/2$, $P_{T_0\sigma}=1/4$, and finally, directly inserting the relevant rates in the limit $\epsilon=0$, $t<T$, we find $P_{T_0T_\pm}=1/2$.  Inserting these results into Eq. \eqref{eq:ki-linear-system} and solving the linear system in terms of the two remaining branching ratios gives:
\begin{eqnarray}
 k_{T_0} = \frac{1}{2-p_+-p_-},\label{eq:k-T0}\\
 k_{T_\pm} = \frac{1}{4}+\frac{p_\pm}{2-p_+-p_-},\label{eq:k-Tpm}
\end{eqnarray}
with branching ratios
\begin{equation}\label{eq:p-plus-minus}
 p_\pm = P_{T_\pm T_0} = \frac{W_{T_{\pm}T_0}}{W_{T_0}}.
\end{equation}
Substituting the cotunneling rates given in Eq. \eqref{eq:cotunneling-rates} into Eqs. \eqref{eq:p-plus-minus}, \eqref{eq:k-T0},  and \eqref{eq:k-Tpm}, and inserting the results for $T>t$ into Eq. \eqref{eq:current-eps-zero} directly gives an expression for the current vs. $B$:
\begin{equation}\label{eq:IvsBfull}
I(B,\epsilon=0) = \frac{4}{3}cT G(B/T)\frac{B/T}{\sinh\left(B/T\right)},
\end{equation}
with
\begin{equation}
 G(x) = 3\frac{2(\cosh x-1)+x\sinh x}{2(\cosh x-1)+5 x\sinh x}.
\end{equation}
Since $G(B/T)$ differs from a constant only at third order in $B/T$, while $I(B,\epsilon=0)$ is exponentially suppressed for $B/T \gtrsim 1$, to a very good approximation we take $G(B/T)\simeq G(0)=1$ leaving the simple expression
\begin{equation}\label{eq:IvsBapprox}
 I(B,\epsilon=0) \simeq \frac{4}{3}cT \frac{B/T}{\sinh\left(B/T\right)};\quad T>t, \Gamma_l\gg W_{\alpha\beta}.
\end{equation}
The approximate expression, Eq. \eqref{eq:IvsBapprox}, is virtually indistinguishable from the full expression given in Eq. \eqref{eq:IvsBfull}, see Fig. \ref{fig:IvsB}.

At zero magnetic field, $B=0$, Eq. \eqref{eq:IvsBapprox} simply gives $I\simeq \bar{n}W_\mathrm{cot}$, where $\bar{n}=4/3$ gives the average number of electrons that pass through the double dot between ``blocking events''.  This number is $4/3$ if three of four $(1,1)$ charge states block current -- in this case, the three spin triplets -- see ref. \onlinecite{Qassemi2009a}. $W_\mathrm{cot} = cT$ is the rate at which any one of the triplets converts to a singlet through a cotunneling process in the limit $T>t$. 

Physically, the current in Fig. \ref{fig:IvsB} falls to zero when $|B|>T$ since the excitation rate out of the ground-state triplet is exponentially suppressed.

\begin{figure}
\includegraphics[width=0.45\textwidth]{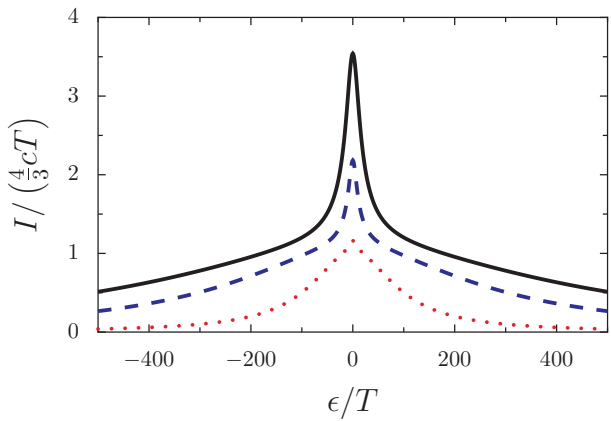}%
\caption{\label{fig:I-vs-eps-vary-t} (Color online) Detuning dependence at $B=0$ from Eq. \eqref{eq:I-vs-eps} for $\alpha=100$ and $t=5T$ (black solid line), $t=3T$ (blue dashed line), and $t=T$ (red dotted line).  The central peak with width $\sim t$ is described by the function $g(\epsilon,t,T)$ and the broad background describes a Lorentzian of width $\delta\epsilon=\alpha t$.}
\end{figure}

\subsubsection{Detuning dependence}
To find the full $\epsilon$-dependence of the current, even in the limit $\Gamma_l\gg W_{\alpha\beta}$ we find that it is necessary to include the escape rates from the singlets $W_{S_\pm}$, which can control the resonant-tunneling current at large detuning, where $W_{S_\pm}\sim W_{T_0},W_{T_\pm}$.  

As in the last section, we aim to find a good approximate solution for the current, starting from the linear equations given in Eq. \eqref{eq:ki-linear-system}. In the limit $\Gamma_l\gg W_{\alpha\beta}$, assumed here, the escape rate from the singlets can be taken to be essentially instantaneous except at sufficiently large detuning $|\epsilon| \gg t$, since $W_{\sigma S_\pm}\propto (t/\epsilon)^2\Gamma_R$ for $\epsilon\to \pm \infty$ (the rate $W_{\sigma S_\pm}$ is limited by the overlap of $\left|S_\pm\right>$ with the $(0,2)$ charge state).  It is therefore sufficient to approximate rates by their large-$\epsilon$ forms to determine $k_{S_\pm}$.  We assume that transition rates between the singlets are small compared to the direct escape rate (i.e., $W_{S_-S_+}\sim (t/\epsilon)^2c|\epsilon| \ll \min\left(W_{\sigma S_+},W_{\sigma S_-}\right)\propto (t/\epsilon)^2\Gamma_R$, which simplifies to $c|\epsilon| \ll \Gamma_R$).  Provided this is satisfied, in the limit $|\epsilon|\gg t$, we find that only $k_{S_+}$ ($k_{S_-}$) is relevant for $\epsilon>0$ ($\epsilon<0$), allowing us to introduce a single parameter $k_S=\theta(\epsilon)k_{S_+}+\theta(-\epsilon)k_{S_+}$, where $\theta(x)$ is a Heaviside step function.  At $B=0$, we further find that $k_{T_+}=k_{T_-}=k_{T}$.  The remaining three independent parameters are then given by the linear equations, from Eq. \eqref{eq:ki-linear-system} after inserting all rates:
\begin{eqnarray}
 k_T & = & \frac{1}{4}+\frac{1}{2+g}k_{T_0}+\frac{1}{3+\eta}k_S,\label{eq:kT}\\
 k_{T_0} & = & \frac{1}{4}+\frac{2}{1+g}k_{T}+\frac{1}{3+\eta}k_S,\label{eq:kT0}\\
 k_S & = & \frac{1}{4}+\frac{1}{3}k_{T_0}+k_T,\label{eq:kS}
\end{eqnarray}
with 
\begin{eqnarray}
g=g(\epsilon,t,T)  & = & \frac{\omega_+ F(\omega_-/T)+\omega_- F(-\omega_+/T)}{\sqrt{\epsilon^2+8 t^2}},\\
\omega_\pm & = & \frac{1}{2}\left[\sqrt{\epsilon^2+8 t^2} \mp \epsilon\right],\label{eq:omega-pm}
\end{eqnarray}
and
\begin{equation}\label{eq:delta-definition}
 \eta = \frac{\sum_\sigma W_{\sigma S}}{W_\mathrm{cot}}\simeq \frac{2\Gamma_R}{cT}\left(\frac{t}{\epsilon}\right)^2 = \frac{16}{3}\left(\frac{\delta\epsilon}{\epsilon}\right)^2,\quad |\epsilon|\gg t.
\end{equation}
Here we have introduced a new energy scale $\delta\epsilon$, giving the value of the detuning at which the inelastic cotunneling rates are comparable to the escape rate from the double dot:
\begin{equation}
\delta\epsilon  =  \alpha t;\quad \alpha = \sqrt{3\Gamma_R/8cT}.\label{eq:delta-eps}
\end{equation}

\begin{figure}
\includegraphics[width=0.45\textwidth]{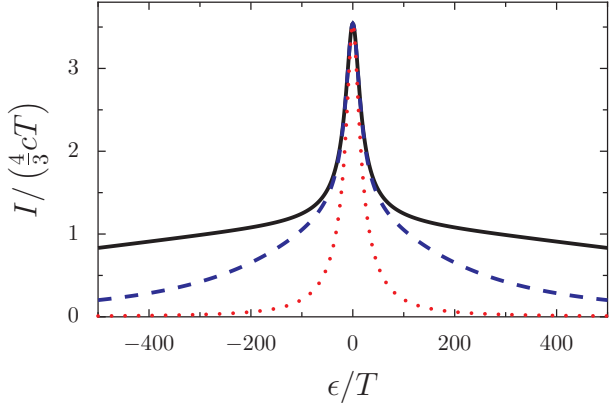}%
\caption{\label{fig:I-vs-eps-vary-alpha} (Color online) Detuning dependence at $B=0$ from Eq. \eqref{eq:I-vs-eps} for $t=5T$ and $\alpha=200$  (black solid line), $\alpha=50$ (blue dashed line), and $\alpha=10$ (red dotted line).}
\end{figure}

As the detuning $\epsilon\to \pm \infty$ is increased, the escape rate from the relevant singlet state $W_{\sigma S_\pm}\propto (t/\epsilon)^2\Gamma_R$ decreases until it becomes smaller than the cotunneling rates $\sim cT$.  In the extreme limit, $\eta=0$ and $g=1$,  Eqs. \eqref{eq:kT}, \eqref{eq:kT0}, and \eqref{eq:kS} only have the singular solution $k_T=k_{T_0}=k_S=\infty$, reflecting the fact that each state is visited an infinite number of times; the system becomes `stuck' in loops, as depicted between states $\left|\alpha\right>$ and $\left|\beta\right>$ in Fig. \ref{fig:transport-fig}. To arrive at the leading finite corrections in the limit of large detuning $\epsilon \gg \delta\epsilon$ (equivalently, $\eta \ll 1$), we set $g=1$, and determine the leading asymptotic solution of Eqs. \eqref{eq:kT}, \eqref{eq:kT0}, and \eqref{eq:kS} for small $\eta$, giving
\begin{eqnarray}
 k_T & =&  k_{T_+}=k_{T_-} = \frac{2}{\eta} + \mathrm{const.},\quad \epsilon\gg \delta\epsilon,\\
k_{T_0} & \simeq & k_{S} = \frac{3}{\eta} + \mathrm{const.},\quad \epsilon \gg \delta\epsilon.
\end{eqnarray}
In the same large-detuning limit, we have the total decay rates (at any finite temperature $T$)
\begin{eqnarray}
W_{T_+} &=& W_{T_-} \simeq 2cT,\quad \epsilon \to \pm \infty,\\
W_{T_0} &\simeq & W_{S_\pm} \simeq 3cT,\quad \epsilon\to \pm \infty.
\end{eqnarray}

In the opposite limit of small detuning ($\epsilon \ll \delta\epsilon$, or equivalently $\eta\gg 1$), Eqs. \eqref{eq:kT} and \eqref{eq:kT0} decouple from Eq. \eqref{eq:kS}.  For $|\epsilon| \lesssim t$ and low temperature $T\lesssim t$, it is necessary to keep the $g$-dependence.  The resulting solutions in this limit are
\begin{eqnarray}
 k_{T_0} & = & \frac{1}{4}+\frac{1}{2g},\quad \epsilon \ll \delta\epsilon,\\
 k_{T} & = & k_{T_\pm} = \frac{1}{4}+\frac{1}{4g},\quad \epsilon \ll \delta\epsilon,
\end{eqnarray}
with corresponding rates given by
\begin{eqnarray}
 W_{T_\pm} = cT\left(1+g\right),\\
W_{T_0} = cT\left(2+g\right).
\end{eqnarray}

Combining the above results gives
\begin{eqnarray}
 \sum_j \frac{k_j}{W_j} &\simeq & \frac{4}{cT}\frac{1}{\eta}=\frac{3}{4cT}\left(\frac{\epsilon}{\delta\epsilon}\right)^2,\quad \epsilon \gg \delta\epsilon,\label{eq:Sumklargeeps}\\
 \sum_j \frac{k_j}{W_j} &\simeq & \frac{3}{4cT}g^{-1}(\epsilon,t,T),\quad \epsilon \ll \delta\epsilon.\label{eq:Sumksmalleps}
\end{eqnarray}
Since the result in Eq. \eqref{eq:Sumklargeeps} vanishes for $\epsilon\ll \delta\epsilon$, but dominates over Eq. \eqref{eq:Sumksmalleps} for $\epsilon\gg \delta\epsilon$, we can simply add the two results to find the appropriate denominator for the current, Eq. \eqref{eq:current-k-formula}, giving an expression that closely approximates the current everywhere except possibly in a small region around $\epsilon \sim \delta\epsilon$.

The resulting lineshape for the leakage current as a function of detuning $\epsilon$ is
\begin{equation}
 I(B=0,\epsilon)  \simeq  \frac{\frac{4}{3}cT}{g^{-1}(\epsilon,t,T)+\left(\epsilon/\delta\epsilon\right)^2},\quad \Gamma_l\gg W_{\alpha\beta}.\label{eq:I-vs-eps}
\end{equation}
In general, the leakage current lineshape may be dominated by the function $g(\epsilon,t,T)$, due to escape from the triplets at small detuning $\epsilon\lesssim t$, and by a broad Lorentzian with width $\delta\epsilon$, limited by escape from the singlets at large detuning, $\epsilon > t$.  Eq. \eqref{eq:I-vs-eps} is plotted in Fig. \ref{fig:I-vs-eps-vary-t} for various values of the tunnel coupling $t$, demonstrating the crossover from a narrow central peak dominated by triplet relaxation to a broad Lorentzian background, when each of the states is visited many times before an electron escapes the double dot.  The balance between broad Lorentzian and peaked resonant tunneling can be tuned with the ratio of escape rate $\sim \Gamma_R$ to cotunneling rate $\sim cT$, controlled by the parameter $\alpha$.  The evolution of the current vs. detuning as $\alpha$ is varied is shown in Fig. \ref{fig:I-vs-eps-vary-alpha}.

\begin{figure}
\includegraphics[width=0.45\textwidth]{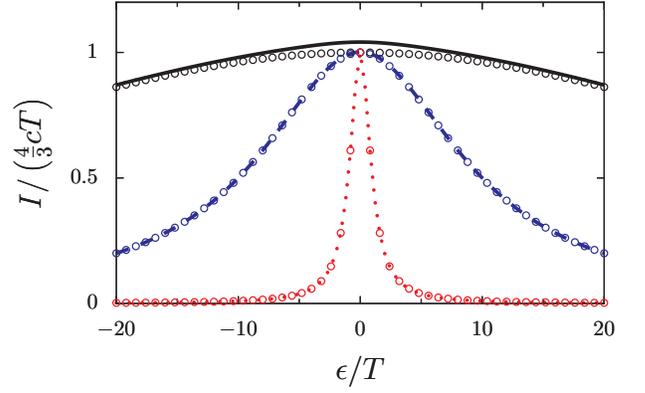}%
\caption{\label{fig:I-vs-eps-vary-t-small-t} (Color online) Detuning dependence at $B=0$ from Eq. \eqref{eq:I-vs-eps} for $\alpha=100$ and $t=0.5T$ (black solid line), $t=0.1T$ (blue dashed line), and $t=0.01T$ (red dotted line).  Open circles give the equivalent curves from Eq. \eqref{eq:I-vs-eps-high-T}, valid in the limit $t\lesssim T$.}
\end{figure}

The lineshape given in Eq. \eqref{eq:I-vs-eps} simplifies considerably in the high-temperature limit $T>t$, in which case the function $g(\epsilon,t,T)\simeq 1$, leaving a simple Lorentzian:
\begin{equation}\label{eq:I-vs-eps-high-T}
 I(B=0,\epsilon) \simeq \frac{\frac{4}{3}cT}{1+\left(\epsilon/\delta\epsilon\right)^2},\quad  T>t, \Gamma_l\gg W_{\alpha\beta}.
\end{equation}
Eq. \eqref{eq:I-vs-eps-high-T} is plotted in Fig. \ref{fig:I-vs-eps-vary-t-small-t} and compared with the full expression given in Eq. \eqref{eq:I-vs-eps} in the relevant high-temperature limit.  

Eq. \eqref{eq:I-vs-eps-high-T} is consistent with recent experiments on silicon double quantum dots.\cite{Lai2010a}

\subsection{Leakage current in the strong-dephasing limit}\label{sec:strong-dephasing}

\begin{figure}
\includegraphics[width=0.45\textwidth]{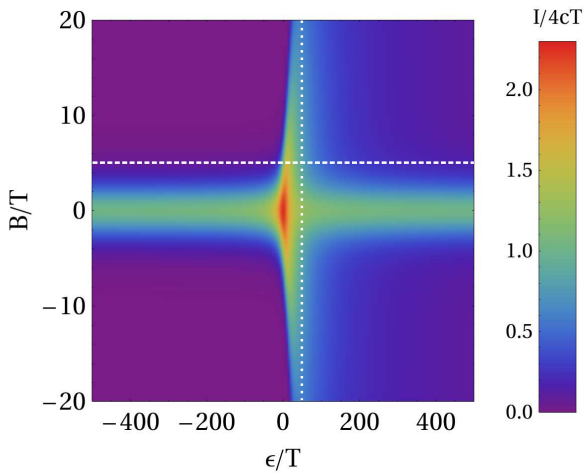}%
\caption{\label{fig:IvsB-density-plot} (Color online) Map of the leakage current vs. detuning, $\epsilon$, and magnetic field, $B$, in the strong-dephasing limit, $W_{T_0}\gg W_{T_\pm}$ from Eqs. \eqref{eq:Current-Dephasing} and \eqref{eq:TpmEscapeRates}.  A tunnel coupling $t=5T$ was chosen to generate this plot.  Cuts vs. $B$ at finite $\epsilon$ (dotted line) and vs. $\epsilon$ at finite $B$ (dashed line) are shown in Figs. \ref{fig:IvsB-cut} and \ref{fig:Ivseps-cut} below, respectively.}
\end{figure}

\begin{figure}
\includegraphics[width=0.45\textwidth]{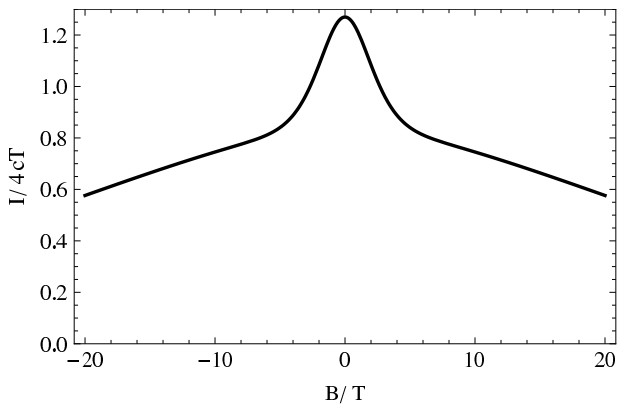}%
\caption{\label{fig:IvsB-cut} Cut of magnetic field dependence at finite detuning $\epsilon/T=50$ along the dotted vertical line in Fig. \ref{fig:IvsB-density-plot}.  At finite positive detuning $\epsilon$, the magnetic-field dependence shows a central peak with width set by $\sim T$ due to inelastic escape processes involving the triplet $T_0$ and excited-state singlet $S_+$, followed by a long slow decay at larger $B$, with a width $B\sim\epsilon$, after which the ground state becomes a spin triplet.}
\end{figure}

\begin{figure}
\includegraphics[width=0.45\textwidth]{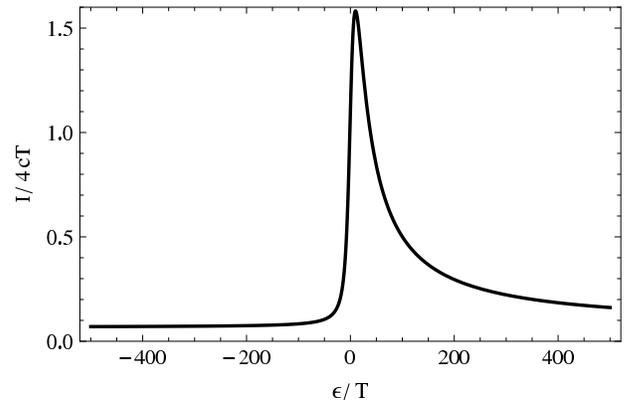}%
\caption{\label{fig:Ivseps-cut} Cut of the detuning dependence at fixed finite magnetic field $B/T = 5$, along the dashed horizontal line in Fig. \ref{fig:IvsB-density-plot}.  At finite magnetic field, the detuning-dependent leakage current due to inelastic cotunneling is asymmetric in $\epsilon$, since the ground state has spin-singlet character for $\epsilon>0$, but spin-triplet character for $\epsilon<0$.}
\end{figure}

An especially simple and ubiquitous limit occurs when there is a strong local dephasing process\footnote{A spin dephasing process that acts locally on the right and left spins of a double quantum dot will convert the coherent triplet $\ket{T_0}=\left(\ket{\uparrow\downarrow}+\ket{\downarrow\uparrow}\right)/\sqrt{2}$ to an incoherent mixture of $\ket{\uparrow\downarrow}$ and $\ket{\downarrow\uparrow}$, both of which have a finite overlap with the singlets $\ket{S_\pm}$, and hence, a finite transition rate to $(0,1)$ charge states via dot-lead tunneling.} or a magnetic field gradient, allowing rapid escape for the spin-unpolarized triplet state $\left|T_0\right>$ (i.e., $W_{T_0}\simeq W_{\sigma T_0}\gg W_{\alpha\beta}$).  The advantage of this limit is the absence of closed `loops' that complicated the analysis in Sec. \ref{sec:no-dephasing}. 

At sufficiently small detuning $\epsilon<\delta\epsilon$, the rate of transition from one of the singlet states to a triplet is small compared to the singlet escape rate, $W_{\alpha S_\pm}\ll W_{\sigma S_\pm}$.  Consequently, the spin-polarized triplets $\left|T_\pm\right>$ are visited at most once in each transport cycle, giving $k_{T_\pm}=1/4$, since each of the four spin states in the $(1,1)$ charge configuration has equal probability of being occupied during a transport cycle.  This gives immediately, from Eq. \eqref{eq:current-k-formula},
\begin{multline}\label{eq:Current-Dephasing}
 I \simeq  \frac{4}{W_{T_+}^{-1}+W_{T_-}^{-1}},\\
 \Gamma_l \gg W_{\alpha\beta},\quad W_{T_0}\gg W_{T_\pm},\quad W_{\sigma S_\pm}\gg W_{\alpha S_\pm}.
\end{multline}
It is important to emphasize the generality of the simple expression given in Eq. \eqref{eq:Current-Dephasing}.  In particular, this expression is valid for arbitrary spin-flip processes leading to transitions from the spin-polarized triplet states to the singlets or spin-unpolarized triplet $T_0$, e.g. $W_{S_+T_\pm}\ne 0$, $W_{T_0T_\pm}\ne 0$, or direct transitions leading to escape from the double dot, $W_{\sigma T_\pm}\ne 0$.  These processes can be mediated by coupling to nuclear spins, spin-orbit interaction, or any other mechanism.  The existence of a fast local dephasing mechanism (e.g., coupling to nuclear spins) even when spin relaxation rates may be slow is common.  Even if there is no dephasing mechanism, a magnetic field gradient across the double dot is sufficient to reach the limit $W_{T_0}\gg W_{T_\pm}$.\cite{Qassemi2009a} 

When only inelastic cotunneling processes account for the rate $W_{T_\pm}$, the direct transitions carrying an electron out of the double dot vanish $W_{\sigma T_\pm}= 0$ and the remaining contributions can be found directly from Eq. \eqref{eq:cotunneling-rates}:
\begin{multline}\label{eq:TpmEscapeRates}
W_{T_\pm}/cT  = \left(W_{T_0 T_\pm}+W_{S_+ T_\pm}+W_{S_- T_\pm}\right)/cT\\ 
 = F\left(\pm\frac{B}{T}\right)+\frac{\omega_- F\left(\frac{\pm B-\omega_+}{T}\right)+\omega_+F\left(\frac{\pm B+\omega_-}{T}\right)}{\sqrt{\epsilon^2+8t^2}}. 
\end{multline}
We recall that the energies $\omega_\pm$ are definied in Eq. \eqref{eq:omega-pm} and the functions $F(x)$ are defined by Eq. \eqref{eq:Fdefinition}.  Inserting the rates given in Eq. \eqref{eq:TpmEscapeRates} into Eq. \eqref{eq:Current-Dephasing} immediately gives a complete map of the leakage current, with amplitude determined by the high-temperature cotunneling rate $\sim cT$ and all other features determined by only three dimensionless parameters: the magnetic field, detuning, and tunnel coupling, scaled by the temperature: $B/T,\epsilon/T,t/T$.  We show a map of the leakage current as a function of $B/T$ and $\epsilon/T$ in Fig. \ref{fig:IvsB-density-plot} for the case of $t=5T$.

Taking the limit $B=\epsilon=0$, $T\gg t$, we find that the current saturates at a maximum value:
\begin{eqnarray}
 && I(B=0,\epsilon=0) = 4cT = \bar{n}W_\mathrm{cot},\\ && W_\mathrm{cot}=2cT,\quad T\gg t.
\end{eqnarray}
Here, we find the average number of electrons passing through the double dot for each cotunneling event is $\bar{n}=2$, in contrast to the situation in Sec. \ref{sec:no-dephasing}, where $\bar{n}=4/3$ for the same parameter values.  This is due to the fact that now only two out of four of the $(1,1)$ charge states are blocked ($\left|T_\pm\right>$), giving $\bar{n}=4/2=2$.

A cut of $I$ vs. $B$ at finite detuning $\epsilon>0$ is shown in Fig. \ref{fig:IvsB-cut}.  The central peak in this figure has width $\sim T$, due to thermally activated escape from the ground state triplet through the $T_0$ state.  The broad background current falls to zero at $B \simeq \epsilon$, where the ground state becomes a spin-triplet ($\left|T_-\right>[\left|T_+\right>]$ for $B>0$ [$B<0$]).  A similar effect is shown as a function of $\epsilon$ at finite magnetic field $B>T$ in Fig. \ref{fig:Ivseps-cut}. Here, for $\epsilon<0$ the ground state becomes a spin triplet and current is suppressed exponentially in $B/T$, whereas for $\epsilon>0$ the ground state is a spin-singlet and relaxation processes can still lead to escape with a slow rate $\sim 1/\epsilon$ until $\epsilon$ becomes very large.  A detuning asymmetry such as this one is often ascribed to phonon-assisted tunneling, but can result (as it does for the inelastic cotunneling mechanism considered here) from any other mechanism for which excitation is exponentially suppressed relative to relaxation. 

\subsubsection{B-field dependence (low-T limit)}

\begin{figure}
\includegraphics[width=0.45\textwidth]{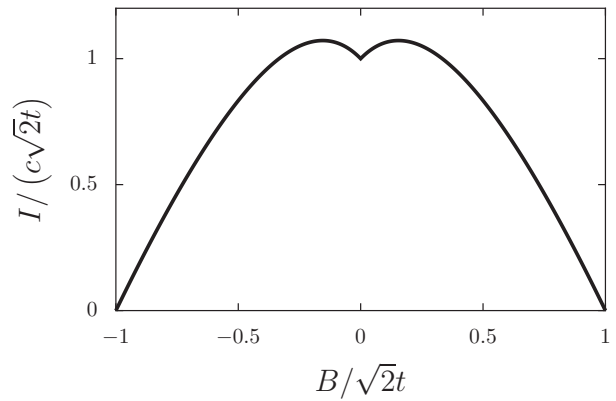}%
\caption{\label{fig:IvsBZeroT} $T=0$ expression for $I$ vs. $B$ (valid for $T<t$).}
\end{figure}

In the low-temperature limit, $T\ll t$, we use the approximation given in Eq. \eqref{eq:F-positive-omega} to find the relevant escape rates from Eq. \eqref{eq:TpmEscapeRates}.  For, e.g., $B>0$ and $\epsilon=0$, these rates are
\begin{eqnarray}
W_{T_+} & = & cB+\frac{c}{2}\left(\sqrt{2}t+B\right)\Theta(\sqrt{2}t+B),\\ 
W_{T_-} & = & \frac{c}{2}\left(\sqrt{2}t-B\right)\Theta(\sqrt{2}t-B).
\end{eqnarray}
Inserting these rates into Eq. \eqref{eq:Current-Dephasing} directly gives the low-temperature expression for the current, previously reported in ref. \onlinecite{Qassemi2009a},
\begin{equation}\label{eq:IvsB-lowT}
I = \frac{c(\sqrt{2}t-|B|)(\sqrt{2}t+3 |B|)}{(\sqrt{2}t+|B|)}\Theta(\sqrt{2}t-|B|),\quad T\ll t.
\end{equation}
Eq. \eqref{eq:IvsB-lowT} is plotted in Fig. \ref{fig:IvsBZeroT}.  The current falls to zero at $|B|=\sqrt{2}t$ when the ground-state triplet falls below the ground-state singlet $\left|S_-\right>$.  At larger $B$, excitation processes are exponentially suppressed and the system becomes locked in the ground-state triplet.  The dip at $B=0$ occurs because relaxation processes from $T_\pm$ to $T_0$ vanish when the levels become degenerate, while at small finite $B$, an additional ``escape route'' is available for the highest-energy triplet through $\left|T_0\right>$.  

From Eq. \eqref{eq:Current-Dephasing}, it is clear that the current will experience a dip at $B=0$ whenever the rates $W_{T_\pm}$ are reduced at $B=0$.  This effect becomes especially pronounced for contributions to $W_{T_\pm}$ from, e.g., the spin-orbit coupling, which must necessarily vanish at $B=0$ due to time-reversal invariance.\cite{Golovach2008a}  This effect due to spin-orbit coupling has been demonstrated in the context of the Pauli spin blockade regime using a phenomenological model that preserves time-reversal, but hybridizes the triplet and singlet states.\cite{Danon2009a}  For a microscopic theory, an additional magnetic-field gradient or local spin dephasing process is likely necessary to arrive at this conclusion in general, since the spin triplet state $\left|T_0\right>$ does not hybridize with the spin singlets at leading order in the spin-orbit coupling.\cite{Golovach2008a}

\section{Conclusions}\label{sec:Conclusions}
We have presented a generic and simple procedure for calculating leakage current through blockaded structures.  Using this generic theory, we found simple analytical expressions for current lineshapes as a function of an applied magnetic field $B$, energy detuning $\epsilon$, and inter-dot tunnel coupling $t$.  These lineshapes fully account for inelastic cotunneling in two limits: weak-dephasing and strong-dephasing.  The results we have found in the weak-dephasing limit are consistent with recent experiments performed on silicon double quantum dots\cite{Lai2010a} and may be applicable to carbon-based double dots, which are also expected to have weak spin-orbit interactions and only weak dephasing.  In the strong-dephasing limit, we have found a simple expression that gives the full two-dimensional map of leakage current vs. $B$ and $\epsilon$ in the presence of inelastic cotunneling.  Finally, we have given a general understanding of zero-field current dips in the limit of strong dephasing.

To simplify the analysis directly associated with the Pauli spin blockade in a double quantum dot, we have neglected orbital and valley degeneracy, which may be relevant in silicon and carbon-based double dots.  Effects of these degeneracies can, however, be included in a systematic and straightforward way using the general methodology outlined in Sec. \ref{sec:blockade-current}.  We leave the details of such an analysis to future study.

\appendix*
\section{Hamiltonian and eigenstates}\label{app:Hamiltonian}

In this Appendix we set the precise definition for the Hamiltonian and isolated double-dot eigenstates.  The starting point is a standard tunneling Hamiltonian for a double quantum dot coupled to leads
\begin{equation}
 H = H_\mathrm{dd}+\sum_l H_l + \sum_l H_{\mathrm{d}l},
\end{equation}
where $H_\mathrm{dd}$ is the Hamiltonian of the double dot, $H_l$ describes Fermi liquid lead $l$, and $H_{\mathrm{d}l}$ gives the tunnel coupling between lead $l$ and dot $l$, with $l=L(R)$ for the left (right) dot/lead, respectively:
\begin{eqnarray}
 H_\mathrm{dd} & = & H_C+H_T+H_Z,\\
 H_l & = & \sum_{k\sigma} \epsilon_{lk\sigma} c_{lk\sigma}^\dagger c_{lk\sigma},\\
 H_{\mathrm{d}l} & = & \sum_{k\sigma} \left(t_lc_{lk\sigma}^\dagger d_{l\sigma}+\mathrm{h.c.}\right).
\end{eqnarray}
Here, $c_{lk\sigma}$ annihilates an electron in lead $l$, orbital state $k$ with spin $\sigma$ having energy $\epsilon_{lk\sigma}$.  The operator $d_{l\sigma}$ annihilates an electron in dot orbital $l$ with spin $\sigma$.  The Coulomb interaction $H_\mathrm{C}$, inter-dot tunneling Hamiltonian $H_\mathrm{T}$ and Zeeman term $H_\mathrm{Z}$ are  
\begin{eqnarray}
 H_C &=& \sum_l \left[\frac{U}{2}n_l(n_l-1)-V_l n_l\right]+U'n_L n_R,\\
 H_T &=& -t\sum_\sigma \left(d_{L\sigma}^\dagger d_{R\sigma}+\mathrm{h.c.}\right),\\
 H_Z &=& \frac{B}{2}\sum_{l}(n_{l\uparrow}-n_{l\downarrow}),
\end{eqnarray}
with number operator defined in the usual way, $n_l = \sum_\sigma n_{l\sigma}$; $n_{l\sigma} = d_{l\sigma}^\dagger d_{l\sigma}$.  In the above expressions, $U$ and $U'$ describe the on-site and nearest-neighbor charging energies, resepectively, in a constant-interaction model, $V_l$ gives the local electrostatic potential for dot orbital $l$, $t$ is the inter-dot tunnel coupling, and $B$ is the applied magnetic field (assumed here to be in-plane so that orbital effects are negligible).

It is convenient to define new energy variables
\begin{eqnarray}
 \epsilon = V_R-V_L-U+U',\\
 \Delta = V_R+V_L-U-U',
\end{eqnarray}
where physically, the energy detuning $\epsilon$ gives the relative energy difference between $(1,1)$ and $(0,2)$ charge configurations and $\Delta$ describes the absolute `depth' of the $(1,1)$ charge configuration. Diagonalizing $H_\mathrm{dd}$ in the space of $(1,1)$, $(0,1)$ and $(0,2)$ charge configurations gives the eigenenergies, assuming a real positive tunnel coupling, $t>0$ (and defining $E_0(\Delta)=-U-\Delta$):
\begin{eqnarray}
 E_\sigma &= & -E_0(\Delta) -\frac{1}{2}(\epsilon-\Delta) +\sigma B/2,\\
 E_{T_{\pm}} &=& -E_0(\Delta) \pm B,\\
 E_{T_0} &=& -E_0(\Delta),\\
 E_{S_{\pm}} &=& -E_0(\Delta) -\frac{1}{2}\left(\epsilon \mp \sqrt{\epsilon^2+8 t^2}\right).
\end{eqnarray}
The associated eigenstates are
\begin{eqnarray}
 \left|\sigma\right> &=& d_{R\sigma}^\dagger \left|0\right>,\\
\left|T_+\right> &=& d_{L\uparrow}^\dagger d_{R\uparrow}^\dagger\left|0\right>,\\
\left|T_-\right> &=& d_{L\downarrow}^\dagger d_{R\downarrow}^\dagger\left|0\right>,\\
\left|T_0\right> &=& \frac{1}{\sqrt{2}}\left(d_{L\uparrow}^\dagger d_{R\downarrow}^\dagger+d_{L\downarrow}^\dagger d_{R\uparrow}^\dagger\right)\left|0\right>,\\
\left|S_\pm\right> & = & \sqrt{C_\pm}\left|S(1,1)\right>\mp \sqrt{C_\mp}\left|S(0,2)\right>.
\end{eqnarray}
where the hybridization of the $\left|S(1,1)\right>$ and $\left|S(0,2)\right>$ singlet states in $\left|S_\pm\right>$ is controlled by the parameters
\begin{equation}
 C_\pm = \frac{\sqrt{\epsilon^2+8 t^2}\pm\epsilon}{2\sqrt{\epsilon^2+8t^2}}.
\end{equation}
The singlets are defined more precisely in terms of creation and annihilation operators by
\begin{eqnarray}
 \left|S(1,1)\right> &=& \frac{1}{\sqrt{2}}\left(d_{L\uparrow}^\dagger d_{R\downarrow}^\dagger-d_{L\downarrow}^\dagger d_{R\uparrow}^\dagger\right)\left|0\right>,\\
\left|S(0,2)\right> &=& d_{R\uparrow}^\dagger d_{R\downarrow}^\dagger\left|0\right>.
\end{eqnarray}

\begin{acknowledgments}
We thank H.~O.~H.~Churchill, A.~Dzurak, N.~S.~Lai, C.~M.~Marcus, A.~Morello, and F.~Zwanenburg for stimulating discussions.  WAC acknowledges funding from the CIFAR JFA, NSERC, FQRNT, and INTRIQ.  FQ acknowledges financial support from NSERC, WIN, and QuantumWorks.
\end{acknowledgments}

\bibliography{leakage}

\end{document}